\documentclass[prd,aps,twocolumn,superscriptaddress]{revtex4-1}

\usepackage{amsmath}
\usepackage{amsfonts}
\usepackage{amssymb}
\usepackage{graphicx}
\usepackage{placeins}
\usepackage[sort&compress]{natbib}
\newcommand{\atKU}{\affiliation{Dept. of Physics and Astronomy, University of Kansas, Lawrence, KS 66045}}
\newcommand{\atOSU}{\affiliation{Dept. of Physics, Center for Cosmology and AstroParticle Physics, The Ohio State University, Columbus, OH 43210}}
\newcommand{\atChiba}{\affiliation{Dept. of Physics, Chiba University, Chiba, Japan}}
\newcommand{\atUW}{\affiliation{Dept. of Physics, University of Wisconsin-Madison, Madison,  WI 53706}}
\newcommand{\atNTU}{\affiliation{Dept. of Physics, Grad. Inst. of Astrophys., Leung Center for Cosmology and Particle Astrophysics, National Taiwan University, Taipei, Taiwan}}
\newcommand{\atUMD}{\affiliation{Dept. of Physics, University of Maryland, College Park, MD 20742}}
\newcommand{\atUC}{\affiliation{Dept. of Physics, Enrico Fermi Institue, Kavli Institute for Cosmological Physics, University of Chicago, Chicago, IL 60637}}
\newcommand{\atUCL}{\affiliation{Dept. of Physics and Astronomy, University College London, London, United Kingdom}}
\newcommand{\atPSUigc}{\affiliation{Center for Multi-Messenger Astrophysics, Institute for Gravitation and the Cosmos, Pennsylvania State University, University Park, PA 16802}}
\newcommand{\atPSUphys}{\affiliation{Dept. of Physics, Pennsylvania State University, University Park, PA 16802}}
\newcommand{\atPSUast}{\affiliation{Dept. of Astronomy and Astrophysics, Pennsylvania State University, University Park, PA 16802}}
\newcommand{\atULB}{\affiliation{Universite Libre de Bruxelles, Science Faculty CP230, B-1050 Brussels, Belgium}}
\newcommand{\atVUB}{\affiliation{Vrije Universiteit Brussel, Brussels, Belgium}}
\newcommand{\atUNL}{\affiliation{Dept. of Physics and Astronomy, University of Nebraska, Lincoln, Nebraska 68588}}
\newcommand{\atWhittier}{\affiliation{Dept. Physics and Astronomy, Whittier College, Whittier, CA 90602}}
\newcommand{\atUD}{\affiliation{Dept. of Physics, University of Delaware, Newark, DE 19716}}
\newcommand{\atNUU}{\affiliation{Dept. of Energy Engineering, National United University, Miaoli, Taiwan}}
\newcommand{\atNPU}{\affiliation{Dept. of Applied Physics, National Pingtung University, Pingtung City, Pingtung County 900393, Taiwan}}
\newcommand{\atDenison}{\affiliation{Dept. of Physics and Astronomy, Denison University, Granville, Ohio 43023}}
\newcommand{\atNDL}{\affiliation{National Nano Device Laboratories, Hsinchu 300, Taiwan}}

\begin{document}


\title{Modeling the refractive index profile n(z) of polar ice for ultra-high energy neutrino experiments} 



 \author{S.~Ali}\atKU
 \author{P.~Allison}\atOSU
 \author{S.~Archambault}\atChiba
 \author{J.J.~Beatty}\atOSU
 \author{D.Z.~Besson}\atKU
 \author{A.~Bishop}\atUW
 \author{P.~Chen}\atNTU
 \author{Y.C.~Chen}\atNTU
 \author{B.A.~Clark}\atUMD
 \author{W.~Clay}\atUC
 \author{A.~Connolly}\atOSU
 \author{K.~Couberly}\atKU
 \author{L.~Cremonesi}\atUCL
 \author{A.~Cummings}\atPSUigc\atPSUphys\atPSUast
 \author{P.~Dasgupta}\atOSU
 \author{R.~Debolt}\atOSU
 \author{S.~de~Kockere}\atVUB
 \author{K.D.~de~Vries}\atVUB
 \author{C.~Deaconu}\atUC
 \author{M.~A.~DuVernois}\atUW
 \author{J.~Flaherty}\atOSU
 \author{E.~Friedman}\atUMD
 \author{R.~Gaior}\atChiba
 \author{P.~Giri}\atUNL
 \author{J.~Hanson}\atWhittier
 \author{N.~Harty}\atUD
 \author{K.D.~Hoffman}\atUMD
 \author{J.J.~Huang}\atNTU
 \author{M.-H.~Huang}\atNTU\atNUU
 \author{K.~Hughes}\atOSU
 \author{A.~Ishihara}\atChiba
 \author{A.~Karle}\atUW
 \author{J.L.~Kelley}\atUW
 \author{K.-C.~Kim}\atUMD
 \author{M.-C.~Kim}\atChiba
 \author{I.~Kravchenko}\atUNL
 \author{R.~Krebs}\atPSUigc\atPSUphys
 \author{C.Y.~Kuo}\atNTU
 \author{K.~Kurusu}\atChiba
 \author{U.A.~Latif}\atVUB
 \author{C.H.~Liu}\atUNL
 \author{T.C.~Liu}\atNTU\atNPU
 \author{W.~Luszczak}\atOSU
 \author{K.~Mase}\atChiba
 \author{M.S.~Muzio}\atPSUigc\atPSUphys\atPSUast
 \author{J.~Nam}\atNTU
 \author{R.J.~Nichol}\atUCL
 \author{A.~Novikov}\atUD
 \author{A.~Nozdrina}\atKU
 \author{E.~Oberla}\atUC
 \author{Y.~Pan}\atUD
 \author{C.~Pfendner}\atDenison
 \author{N.~Punsuebsay}\atUD
 \author{J.~Roth}\atUD
 \author{A.~Salcedo-Gomez}\atOSU
 \author{D.~Seckel}\atUD
 \author{M.F.H.~Seikh}\atKU
 \author{Y.-S.~Shiao}\atNTU\atNDL
 \author{D.~Smith}\atUC
 \author{S.~Toscano}\atULB
 \author{J.~Torres}\atOSU
 \author{J.~Touart}\atUMD
 \author{N.~van~Eijndhoven}\atVUB
 \author{A.~Vieregg}\atUC
 \author{M.-Z.~Wang}\atNTU
 \author{S.-H.~Wang}\atNTU
 \author{S.A.~Wissel}\atPSUigc\atPSUphys\atPSUast
 \author{C.~Xie}\atUCL
 \author{S.~Yoshida}\atChiba
 \author{R.~Young}\atKU
\collaboration{ARA Collaboration}\noaffiliation


\date{\today}


\begin{abstract}
We have developed an {\it in-situ} index of refraction profile n(z) for cold polar ice, using the transit times of radio signals broadcast from an englacial transmitter to 2-5 km distant radio-frequency receivers, deployed at depths up to 200 m. For propagation through a non-uniform medium, Maxwell's equations generally admit two ray propagation solutions from a given transmitter, corresponding to a direct path (D) and a refracted or reflected path (R); the measured D vs. R timing differences (dt(D,R)) are determined by the refractive index profile.   We constrain n(z) near South Pole, where the Askaryan Radio Array (ARA) neutrino observatory is located, by simulating D and R ray paths via ray tracing and comparing simulations to measured dt(D,R) values. We demonstrate that our dt(D,R) timing data strongly favors a glaciologically-motivated three-phase densification model rather than a single exponential scale height model.  Effective volume simulations for a detector of ARA station antenna depths yield a 14 \% increase in neutrino sensitivity over a range of $10^{18} - 10^{21}$ eV using the three-phase model compared to a single exponential.
\end{abstract}


\maketitle

\section{Introduction}
\label{Sec:Intro}

Ultra-High Energy Neutrino (UHEN) experiments such as the Radio Neutrino Observatory in Greenland (RNO-G), the Askaryan Radio Array (ARA), and the proposed IceCube Gen-2 experiment seek to extend the energy window of observed neutrinos beyond the MeV (typical of solar neutrinos) and PeV scales (astrophysical, as observed by IceCube) to $>$PeV (`cosmogenic') energy scales \cite{aguilar2022radio,allison2016performance,aartsen2021icecube}.  Radio detection provides a cost-effective method for constructing detectors with a large sensitive volume, as radio signals propagate farther in ice compared to optical signals.  A major motivation of UHEN experiments is to complement observations of ultra-high energy charged cosmic rays (UHECR) from distant astronomical sources. UHEN are emitted following collisions of UHECR with matter or the Cosmic Microwave Background. Due to their lack of charge and small cross-section, neutrinos are able to propagate, undeflected, through obstacles otherwise opaque to gamma or cosmic rays.  However, these same weakly interacting characteristics  render observation difficult.

Radio propagation in ice, over kilometer-long distance scales, is essential for the radio neutrino experiments. Since, as the neutrino energy increases, the expected neutrino flux sharply decreases,  
radio neutrino experiments must scan over large volumes for long exposure times to achieve measurable neutrino event rates.

Simulations which incorporate models of the complex-valued ice permittivity are used to estimate the sensitivity of UHEN experiments.
 The real part of the complex permittivity dictates the ray path followed by radio signals, from interaction point to receiver, while the imaginary part quantifies the degree to which signal is absorbed in-ice.  Given the non-magnetic nature of ice, the permittivity relates directly to the refractive index profile through $n = \sqrt{\epsilon_r}$ where $\epsilon_r$ is the real part of the ice permittivity. Since the overburden increases with depth, 
 UHEN experiments assume a depth-dependent refractive index ansatz.  For a given receiver, the varying index of refractive profile in the upper $\sim$100-150m generates, by Fermat's Least Time Principle, curved rather than rectilinear ray trajectories. Given a receiver depth, this ray-bending results in a `shadowed' volume within which neutrino interactions will be inaccessible - this loss of sensitivity is increasingly important for large horizontal displacements and shallow receivers.  As the depth of deployed receiver decreases (i.e., receiver gets closer to the surface) the extent of the shadowed zone is an increasingly important determinant of the neutrino interaction volume visible to a given receiver, and therefore the number of detected neutrinos.

The simplest ansatz for the refractive index profile is one that follows a single exponential dependence on depth, as expected for a self-gravitating fluid.  Glaciological studies of ice density as a function of depth, however, suggest that densification occurs in multiple stages \cite{herron1980firn,stevens2020community,salamatin1997bubbly}. Using ARA receiver timing data obtained in response to a pulser lowered into the ice, we have tested a piece-wise function separated into these 3 stages against a simpler one-stage or two-stage exponential model.

\section{Ice Densification and the Refractive Index}
\label{Sec:Ice Densification}

Our model assumes a linear dependence of refractive index on density\cite{robin1969interpretation}, although this relationship has not yet been fully verified in the lab.  According to Sorge's Law \cite{bader1954sorge}, density is constant over time at a given depth, assuming constant snow accumulation and constant temperature.
(In reality, the snow accumulation and temperature conditions are not constant, which leads to density fluctuations; in what follows, we neglect such effects, as well as the possible effect of impurities.) 
In a scale-height model, the densification rate of snow is taken to be proportional to the change in pressure due to the weight of the snow overburden, leading to the exponential form 
\begin{align}
    \rho (z) = \rho_f - b_0 e^{Cz}
\end{align}
where $\rho_f$ is the density of deep ice, $\rho_f - b_0$ is the density of snow at the surface, and $C$ is a proportionality constant prescribed by the densification rate. The density profile $\rho(z)$ can then be translated to n(z) assuming a linear relationship between the two quantities.  

Theoretical and empirical models of ice density as a function of depth, however, prescribe two boundaries that affect the ice densification rate as $\rho(z)$ crosses certain thresholds\cite{salamatin1997bubbly,lipenkov1997bubbly}.  
The (better-studied) first boundary at $\sim$550 kg/m$^3$\cite{herron1980firn} separates the snow and firn regions, and occurs at depths of $\sim$15m and $\sim$20m for Summit and South Polar ice, respectively.\cite{stevens2020community}.  A second boundary separates the firn and bubbly ice regions.  The ice is considered to be fully formed in the third region, but there still exist air pockets within the ice that reduce the overall density relative to $\rho_f$.  These air pockets are crushed under greater pressure as depth increases, asymptotically reaching the $\rho_f\sim$920 kg/m$^3$ density of typical pure, deep ice at T=-30$^\circ$ C\cite{Aguilar:2023udv}.  To account for changing densification rates in the separate regions, the form of Eqn (1) is modified to a piecewise function.
  \begin{align*}
     \rho < \rho_1 : \rho (z) = \rho_f - b_0e^{c_1 z}
\end{align*}
\begin{align*}
     \rho_1 \leq \rho < \rho_2: \rho (z) = \rho_f - b_0 e^{z_1 (c_1 - c_2)} e^{c_2 z}
\end{align*}
\begin{align}
     \rho \geq \rho_2: \rho (z) = \rho_f - b_0 e^{z_1(c1-c2)} e^{z_2(c2-c3)} e^{c_3z}
\end{align}
where $\rho_1$ and $\rho_2$ are density boundary conditions with corresponding depths $z_1$ and $z_2$ that define the regions, $b_0$ is determined by surface density and continuity at each boundary, and $c_1, c_2$ and $c_3$ are constants describing the densification rate in each density region.

\begin{figure}[h]
    \centering
    \includegraphics[width = 8.6cm]{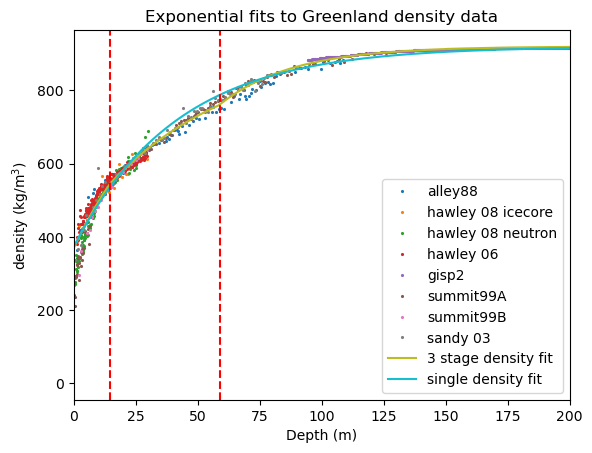}
    \includegraphics[width = 8.6cm]{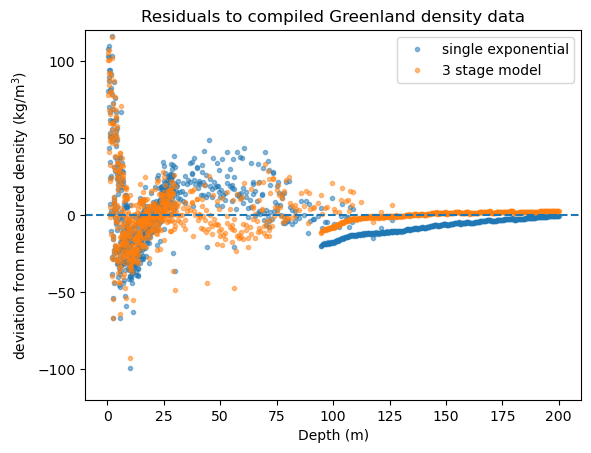}
    \caption{\small Equations (3) (single) and (4) (3 stage) parameterization fits to compilation of Greenland ice density data.  Holding the snow to firn (550 kg/m$^3$) boundary constant, fit gives firn to bubbly ice boundary of 758 kg/m$^3$.  Red dotted lines (14.9m, 58.9m) indicate corresponding boundary depths to density boundaries.  Density data have been taken from \cite{gow1997physical,alley1988ice,hawley2006borehole,hawley2008rapid}.}
    \label{fig:grn_density}
\end{figure}

Figure \ref{fig:grn_density} shows ice density data taken from a number of sites in Greenland.  This density data approaches the asymptotic value and therefore includes measurements well into the expected bubbly ice region. We use a 550 kg/m$^3$ boundary between the snow and firn regions based on previous studies 
\cite{herron1980firn,stevens2020community}, while the boundary between firn and bubbly ice is determined empirically by fitting the $\rho(z)$ data for the inflection point separating the intermediate and highest density regions, using the Levenberg–Marquardt fitting algorithm\cite{gavin2019levenberg}. In this fitting procedure, the surface density, $c_1$, $c_2$, $c_3$, and the firn to bubbly ice boundary are taken as free parameters while the snow to firn boundary and asymptotic density are fixed at 550 kg/m$^3$ and 920 kg/m$^3$ respectively. Each data point from the compiled density data is weighted equally in this fit.  Applying the fitting algorithm gives a 758 kg/m$^3$ boundary between the firn and bubbly ice regions which is used to determine the second depth boundary in Eqn. (6).  Past studies of the snow to bubbly ice density boundary have used a density from 800-830 kg/m$^3$ \cite{herron1980firn,lipenkov1997bubbly}.  For the single exponential, the same fitting algorithm is used and $b_0$ and $c$ in Equation (1) are taken to be free parameters.  This gives a best fit single exponential to the compiled data of 
\begin{align}
    \rho (z) = 920 - 548.5 e^{-0.0241z}
\end{align}
and a 3 stage exponential fit using the form of Equation (2) 
\begin{align*}
     \rho < 550  \mathrm{kg/m}^3 : \rho (z) = 920 - 551.7e^{-0.0262 z}
\end{align*}
\begin{align*}
     550  \mathrm{kg/m}^3 \leq \rho < 758  \mathrm{kg/m}^3: \rho (z) = 920 - 373.4e^{-0.0193(z-14.9)}
\end{align*}
\begin{align}
     \rho \geq  758  \mathrm{kg/m}^3 : \rho (z) = 920 - 159.0e^{-0.0339 (z-58.9)}
\end{align}

\subsection{Calculating refractive index from density}
Ice density is converted to index of refraction n(z) using a linear relationship based on studies of dielectric constant vs. specific gravity \cite{robin1969interpretation}. While there exists variation in the parameters for different sets of data, this linear relation serves as a useful approximation of the dielectric specific gravity relation \cite{kovacs1995situ}. Studies at the Maudheim and McMurdo ice shelfs as well as studies of snow in Canada and Japan have supported a relation.\cite{schytt1958snow,evans1965dielectric,cumming1952dielectric}

\begin{align}
    n(z) = 1 + A \rho, 
\end{align}

where $\rho$ is the pure ice specific-gravity and $A$ is a proportionality constant of units cm$^3$/g, which can be estimated from the constraint that deep, bulk ice has a refractive index corresponding to the measured value at a given site. At Summit Station, Greenland, n=1.778$\pm$0.006 \cite{Aguilar:2023udv, welling2023precision} for bulk ice. Different fits to McMurdo ice shelf data give $A$ values ranging from 0.840-0.858. The $A$ value of 0.845 determined from the most recent McMurdo ice shelf study \cite{kovacs1995situ} also matches with the conversion of the asymptotic ice density 920 kg/m$^3$ to the value n=1.778.  For this reason, we will be using 0.845 as the conversion factor from SPICE core density to n(z) (see Section VI).  Application of Eqn. (5) translates the exponentially-assumed density data into a two-parameter exponential n(z) profile, customarily written as:

\begin{align}
    n(z) = 1.778 - B_0 e^{Cz}.
\end{align}

where $B_0$ is $A \times b_0$ from Eqn. (5) and (1) respectively and $C$ is constant from Eqn. (1).  Applying the same density to refractive index conversion to Eqn. (2) gives a 3 stage refractive index model of the form
  \begin{align*}
     z < z_1 : n (z) = 1.778 - B_0e^{c_1 z}
\end{align*}
\begin{align*}
    z_1 \leq z < z_2: n (z) = 1.778 - B_0 e^{z_1 (c_1 - c_2)} e^{c_2 z}
\end{align*}
\begin{align}
     z \geq z_2: n (z) = 1.778 - B_0 e^{z_1(c_1-c_2)} e^{z_2(c_2-c_3)} e^{c_3z}
\end{align}
where $c_1$, $c_2$, $c_3$ are exponential parameters to be determined by the fitting the model to timing data.  


\subsection{Experimental Layout}
\begin{figure}[h]
    \centering
    \includegraphics[width = 8.6cm]{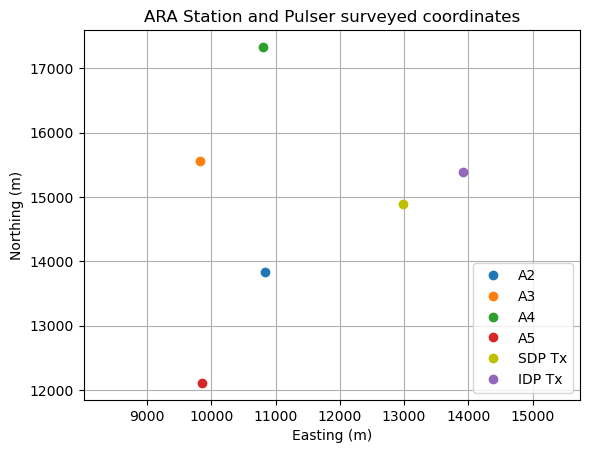}
    \caption{\small Experimental layout. Surveyed coordinates in Easting, Northing for ARA stations A2, A3, A4 and A5 for which timing data from the SPICE (SDP) and IC1S (IDP) are analyzed. The SDP transmitter was lowered from the surface to a depth of 1700 m, and the IDP transmitter was at a fixed depth of approximately 1400 m.}
    \label{fig:geo}
\end{figure}

Analyzed ARA data discussed below are drawn from multiple and redundant combinations of transmitters and receivers, spanning in-ice transmitter depths of up to 1.5 km and horizontal trajectories of up to 5 km over a wide range of incidence angles. A plane view of the relevant radio-frequency instrumentation in the vicinity of South Pole is presented in Figure \ref{fig:geo}, showing the receiver stations relative to the radio-frequency transmitters.

\section{Deep Pulser Data}

\begin{figure}[ht]
    \centering
    \includegraphics[width = 8.6cm]{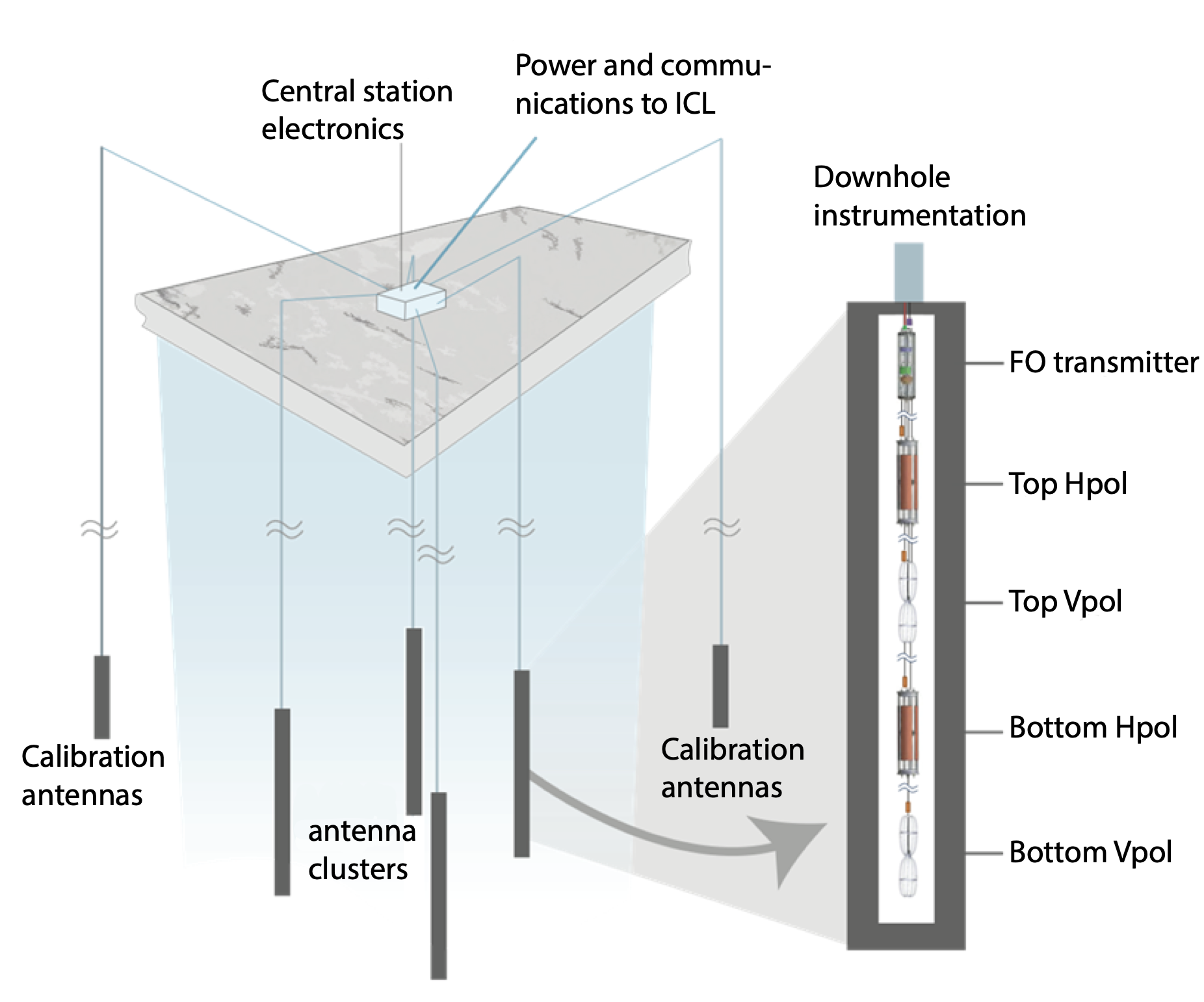}
    \caption{\small ARA station layout, comprising 16 antennas (8 Hpol and 8 Vpol) ranging from 170-200m depths.}
    \label{fig:ARAstation}
\end{figure}

A typical ARA station is shown in Figure \ref{fig:ARAstation}; ARA consists of 5 such stations, denoted as `A1'-`A5'.  Each station consists of 4 boreholes at the vertices of a horizontal square, with an inter-string lateral separation of order 20 meters; each borehole contains a  vertical string of radio receiver antennas.  Each string consists of 2 horizontally polarized (Hpol) and 2 vertically polarized (Vpol) antennas.  This allows the stations to record both Hpol and Vpol signals, as well as infer the polarization of a signal based on the relative amplitudes of the signals registered on the Hpol vs. Vpol antennas.  The array of antennas allows for the reconstruction of a source location based on the relative timings for signals received by multiple antennas. For a typical station, the top Hpol to bottom Vpol antennas are deployed at depths ranging from 170-200m.  Nearby englacial pulsers are used to calibrate channel positions and accuracy using relative timing differences, assuming some n(z) profile.

In our case, we are interested in not only the direct (D) signal arrival time, but also the refracted/reflected (R) signal arrival time. In an effort to reduce systematic errors, a combination of two timing methods are used. 
 In both cases a waveform similar to that seen in Figure 4 is split into its direct and refracted signals. The timing difference is calculated by finding the maximum cross correlation of the two signals.  This is referenced with a method that takes the leading edges of the two signals by taking the time where the amplitude crosses a threshold of 1/3 the maximum signal amplitude.  The average absolute discrepancy of these two methods is 2.49 ns with a standard deviation of 1.29 ns across the deep pulser data.  We calculate the timing difference dt(D,R) by taking the average value of these two timing methods.

\begin{figure}[ht]
    \centering
    \includegraphics[width = 8.6cm]{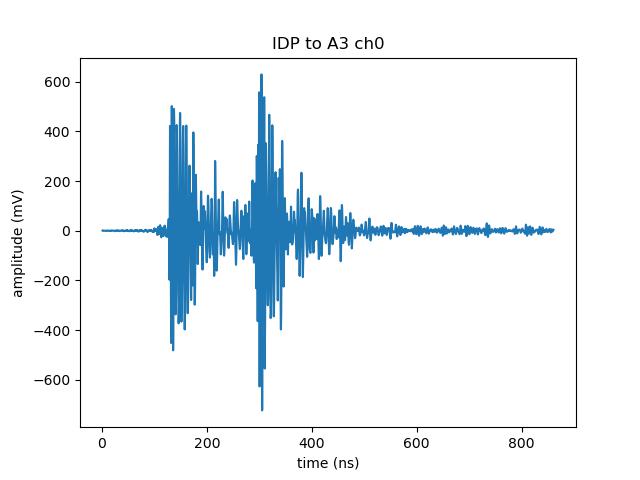}
    \caption{\small  Example stationary IceCube deep pulser (IDP) averaged waveform at 1400m depth, illustrating the direct (first) and refracted (second) signals measured in ARA station A3.}
    \label{fig:waveform}
\end{figure}

The dt(D,R) timing difference offers several advantages in calibration.  dt(D,R) is immune to data acquisition signal transmission (aka `cable delay') uncertainties since the cable delays are equal for the D and R signals on a single channel waveform capture.  The dt(D,R) method is also particularly useful for testing n(z) models due to the differences between the D and R optical paths (see Figure \ref{fig:raytrace_path}).  The R ray path travels through ice well above the receiver antenna which allows for testing a greater range of depths for n(z).  By contrast, nearby calibration pulsers only test n(z) over the limited depth range of the deployed ARA antennas. From a science perspective, since dt(D,R) (approximately) linearly depends on the range to source location, it can also be used as a powerful constraint in neutrino reconstruction, independent of conventional interferometric techniques to locate interaction points \cite{latif2020towards}.

Figure \ref{fig:waveform} shows a 128-event average waveform obtained from a stationary deep pulser deployed in one of the iceholes drilled for the IceCube experiment (referred to as `IceCube Deep Pulser', or `IDP'; z=1400 m). 
The deep pulser is horizontally displaced from the ARA stations by typical distances of 1-5 km, depending on the ARA receiver station. This IDP source signal was observed in all 5 ARA stations.

Additional data from a deep pulser lowered into the South Pole Ice Core (SPICE) borehole in the 2018-2019 seasons is also useful in discriminating between refractive index models. The SPICE borehole was drilled to 1751m over the 2014-2015 and 2015-2016 austral seasons with the purpose of collecting data to determine changes in atmospheric chemistry, climate, and biogeochemistry since the most recent [40 ka] glacial-interglacial cycle \cite{casey20141500}.  Proximal to the ARA stations, SPICE offers both density data and the opportunity for deep pulser timing measurements. 

 The South Pole UNiversity of Kansas Pressure Vessel Antenna, or SPUNK PVA transmitter antenna was deployed into the SPICE borehole located 2-4 km from stations A2-A5 \cite{allison2020long}.  We refer to this pulser as the SDP.  The SDP emitted 1 pulse per second (pps) signals to a depth of approximately 1700m.  Stations began receiving double pulse (D,R) signals once the transmitter emerged from the shadow zone (around a depth of 500--700m, depending on the receiver station) resulting in a dataset of dt(D,R) pulses recorded to z=1300m.  A more detailed description of the SDP pulsing runs and received signals can be found in the literature \cite{allison2020long}.  Since the D and R ray paths differ over this range, this dataset provides a check on the consistency of n(z) models.

\section{Simulated Deep Pulser Signals}

\begin{figure}[ht]
    \centering
    \includegraphics[width = 8.6cm]{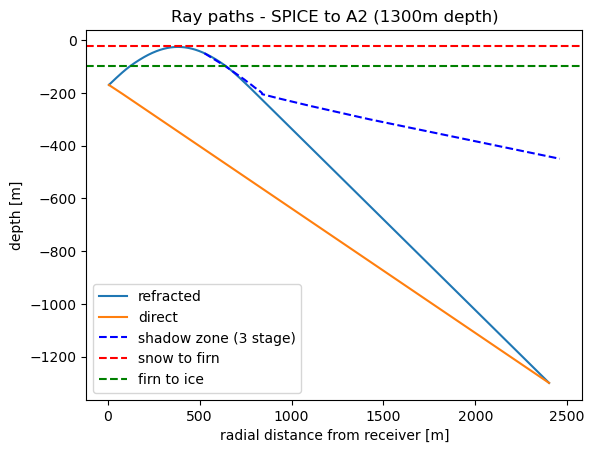}
    \caption{\small Sample simulated (using Eqn. (6)) direct and refracted ray paths from a 1300m deep pulser in the SPICE borehole to channel 0 of ARA staton A2.  Horizontal dotted lines refer to the boundaries between regions of the 3 stage n(z) model.  Shadow zone refers to the area above and to the right of the dashed blue line where pulses cannot be detected.}
    \label{fig:raytrace_path}
\end{figure}

\begin{figure}[ht]
    \centering
    \includegraphics[width = 8.6cm]{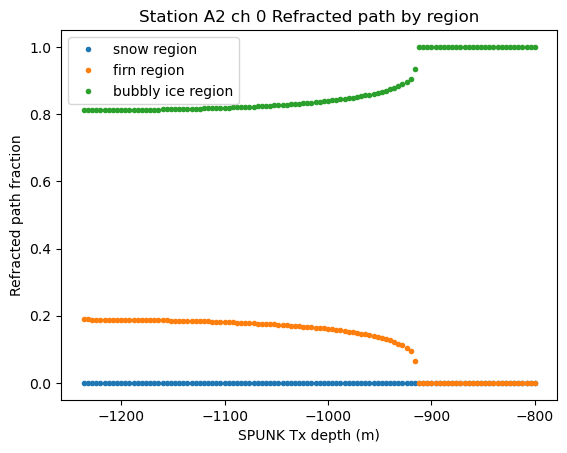}
    \caption{\small Fraction of simulated (using Eqn. (6)) refracted ray path in the snow, firn, and bubbly ice regions from SPICE to station A2 channel 0 (170m depth) for various SPICE transmitter depths.}
    \label{fig:raytrace_regions}
\end{figure}

Ray paths, originating at either the IDP or SDP transmitters, are simulated using the numerical ray tracer {\tt RadioPropa} \cite{Winchen_2019}, which calculates the two possible solutions (D and R), given source and receiver positions and some n(z) model.  The travel time is calculated for each of the ray paths, and dt(D,R) is calculated directly from the difference in travel time for the two paths.  Figure \ref{fig:raytrace_path} shows the simulated ray paths from a 1300m deep source to an ARA antenna.  The D ray path travels only through the bubbly ice region while the R path travels through both the bubbly ice and firn regions of the 3 stage model.  Figure \ref{fig:raytrace_regions} shows the relative path lengths in each region of the simulated R ray path over a range of SDP depths.  As the apex of the R path approaches the firn to bubbly ice boundary the discontinuous gradient of the piecewise function results in a lack of simulated D and R solutions for this transmitter depth.  This results in the gap in simulated values seen in Figure \ref{fig:raytrace_regions}.

\:

The ray tracing code {\tt RadioPropa} is incorporated into {\tt NuRadioMC}, a complete Monte Carlo simulation suite that can be used to simulate signal waveforms and determine the neutrino detection volume for a given radio neutrino detector configuration \cite{glaser2020nuradiomc}.  
In addition to relative signal arrival times, simulated waveforms can be compared to the measured waveforms in terms of the relative amplitudes of the D and R signals as well as frequency content. In what follows we consider only timing information, although amplitude analysis is currently underway.  


\section{Snow Accumulation Impact on Refractive Index Profile}
Before comparing our simulated dt(D,R) to data, it is important to consider possible effects of snow accumulation with time, since ARA receiver depths were recorded at the time of initial deployment (2011--2013 for stations A1--A3), while SDP pulser data were recorded in December, 2018.
Yearly snowfall at the South Pole increases the overall depths, relative to the surface, of antennas buried in prior years.  
A study at South Pole from 1983-2010 using a near-surface snow stake field near Amundsen-Scott station \cite{lazzara2012fifty} found an average annual snow accumulation rate of 274.85 mm per year with a (non statistically-significant) downward trend of $-2.8 \pm 6.7$ mm in annual accumulation over that time. However, over multiple years the increment in depth for sub-surface antennas is smaller than the annual accumulation at the surface due to the densification of snow to ice over time. We calculate this effect by fitting the shallowest region of Eqn. 2 to density data taken in the SPICE borehole using the procedure discussed in Section VI.  For the region $\rho < 550$ kg/m$^3$, this fit becomes that of Eqn. (2) with parameters $p_f = 920$ kg/m$^3$, $b_0=532.5$ kg/m$^3$, and C=0.0147. Taking surface accumulation density to be depth z = 0 m, we calculate the depth increase over multiple years by integrating over this equation to match the total mass of accumulated snow based on the average accumulation at density z=0.  

\begin{figure}[ht!]
    \centering
    \includegraphics[width = 8.6cm]{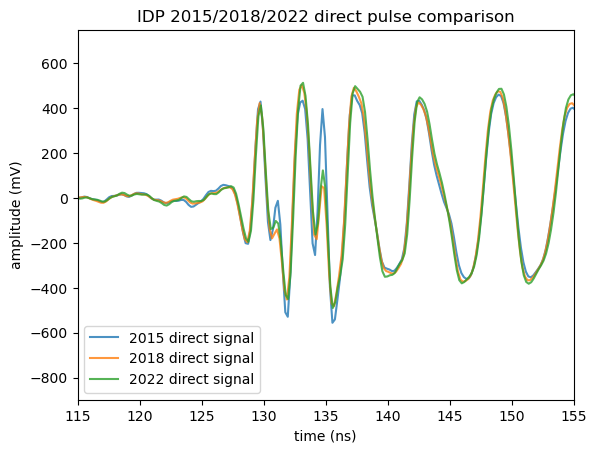}    \includegraphics[width = 8.6cm]{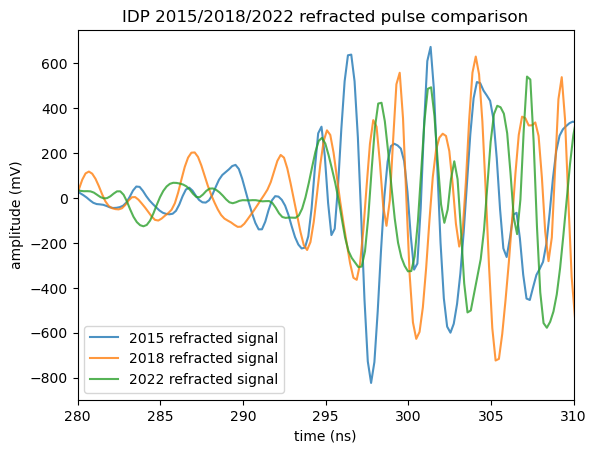}
    \caption{\small Comparison of 2015 vs. 2018 vs. 2022 IceCube Deep Pulser signals, showing direct (top) and refracted (bottom) signals. For the receiver channel considered (at a depth of 175 meters, and therefore close to the asymptotic ice density regime), the direct signals are remarkably consistent with each other (indicating very little aging or hole closure effects), while the refracted signals, which sample shallower snow and are more sensitive to refractive index changes, show evident differences.}
    \label{fig:A3IDP151822}
\end{figure}

ARA Station A3 was deployed during the 2012-2013 season, two years after the IceCube deep pulsers were deployed.  The SPICE borehole data were taken in the 2018-2019 season, corresponding to 6 years of snow accumulation since the initial antenna depths were recorded.  Applying 6 years of accumulation using the 274.85 mm surface accumulation average implies an estimated 1.57 m of snow accumulation over this period.  We can also estimate the depth change at a given antenna position by comparing dt(D,R) times for signals from one of the two IDP transmitters broadcast to ARA station A3, registered over a multi-year time period. Figure \ref{fig:A3IDP151822} overlays the direct and refracted signals for data recorded over an 7 year timespan. In contrast to the nearly-constant direct signals, the refracted signals, with much shallower trajectories, show clear discrepancies. The constancy of the direct signals over time indicate that the variations in the observed refracted waveforms are not the result of, e.g., hole closure effects or some other effect leading to a change in the antenna response over this timescale. 

\begin{figure}[ht!]
    \centering
    \includegraphics[width = 8.6cm]{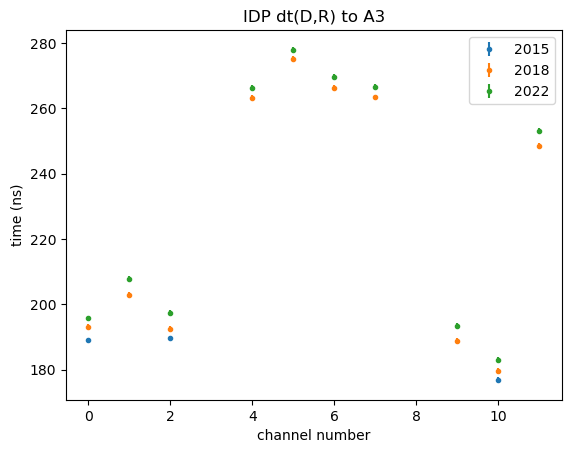}
    \includegraphics[width = 8.6cm]{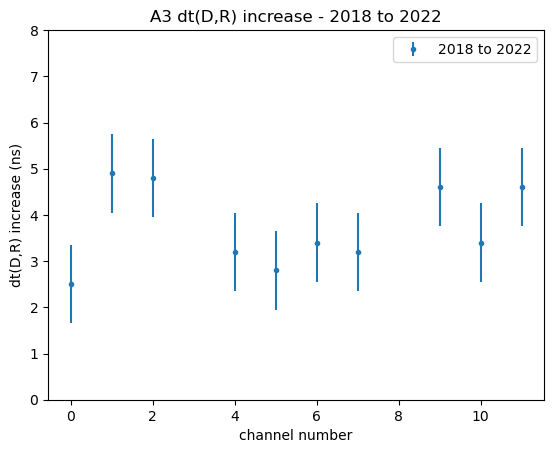}
    \caption{\small Measured dt(D,R) from IDP to A3 antennas for 3 seasons (top) and relative increment in dt(D,R) (bottom); as expected, dt(D,R) increases for later seasons in accordance with increased antenna depth due to snow accumulation.  The highest SNR channels 0,2,10 provide dt(D,R) data across all 3 seasons while remaining channels include dt(D,R) data only for the 2018 and 2022 seasons. Statistical error bars of 0.85 ns from the standard deviation of channel dt(D,R) increases included. }
    \label{fig:A3snow_acc}
\end{figure}

Combining data from several years and selecting those channels with the highest Signal-to-Noise Ratio, Figure \ref{fig:A3snow_acc} shows the extracted dt(D,R) times from the IC22S IDP to station A3 for 2015, 2018, and 2022 data.   
Averaged over all channels, the measured dt(D,R) from IC22S to A3 shows an increase of 6.9 ns (0.99 ns per yr) from 2015-2022 and an increase of 3.7 ns (0.93 ns per yr) from 2018-2022.  Using {\tt RadioPropa}'s numerical ray tracer we simulate the dt(D,R) times from IC22S to A3 and increase the depths of both the pulser and the station antennas in 0.01m increments until the simulated dt(D,R) matches the measured dt(D,R).  For this process, the 3 stage model parameters in Table 1 are used.  The measured dt(D,R) increases correspond to snow accumulations of 1.72 m (0.245 m per yr) from 2015-2022 and 0.936 m (0.234 m per yr) from 2018-2022 using the 3 stage model (5) to simulate dt(D,R). In comparison, the 1983-2010 average annual surface snow accumulation would predict 1.82 m (0.26 m per yr) from 2015-2022 and 1.07 m (0.27 m per yr) from 2018-2022 when applying multiple years of accumulation to sub surface antennas.  The 2015-2022 and 2018-2022 IC22S data imply an average annual snow accumulation of $0.24 \pm 0.09$ m (statistical errors only).  We use this measured snow accumulation rate to correct station antenna depths recorded during deployment to the depths at the time of SPICE borehole data collection.  The IDP is a stationary pulser whose position is assumed constant relative to the receiver channels in these timing comparisons.  As the channels are all localized to a single station, the expected snow accumulation is equal among the channels.  As seen in Figure 8, the measured increase in dt(D,R) is not equal among all channels.  The standard deviation of the increase in dt(D,R) across channels is 0.85 ns. This is included in the timing uncertainty for Figures 12, 13 , and 14 of Section VI.

\section{Ice model results}

\begin{figure}[ht!]
    \centering
    \includegraphics[width = 8.6cm]{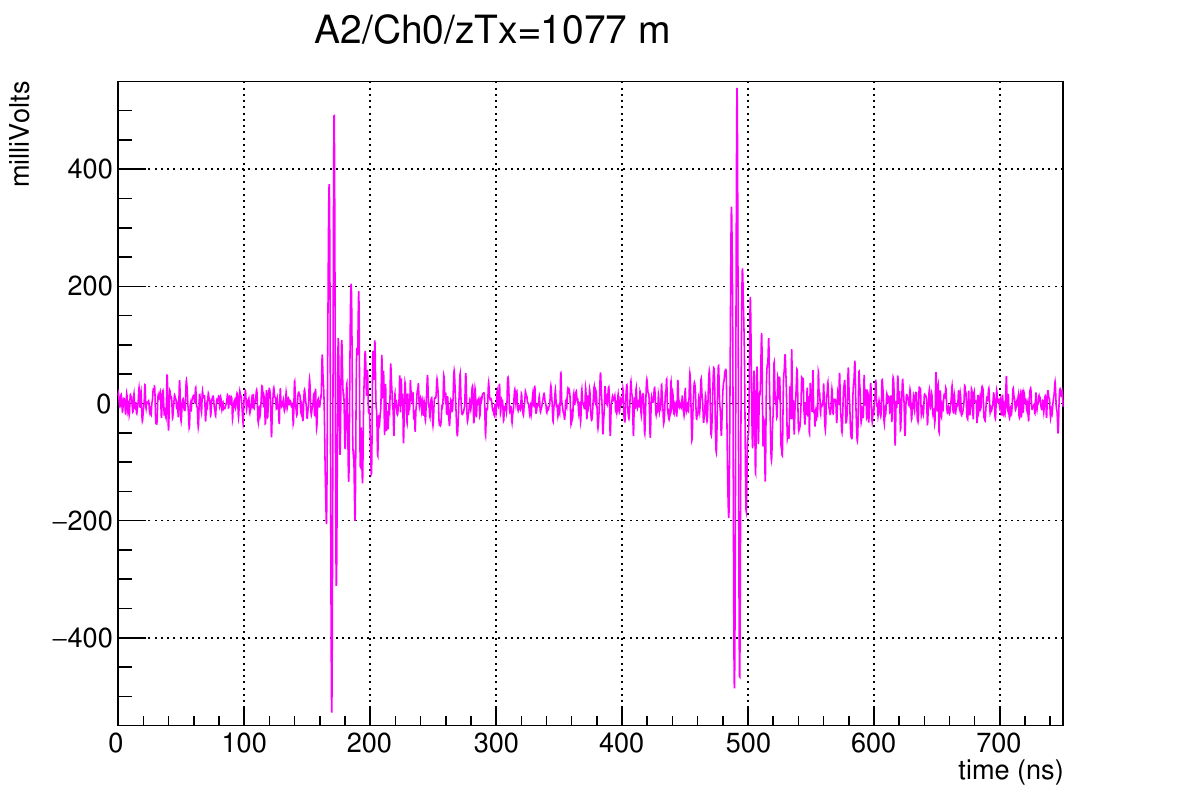}
    \caption{\small Typical waveform recorded during SDP transmitter run; direct and refracted pulses are evident in the Figure.}
    \label{fig:SPICEwf}
\end{figure}

By lowering a pulser into the SPICE borehole, dt(D,R) data were collected for a variety of ray paths. To ensure the transmitter is well beyond the shadow zone, we measure dt(D,R) over the depth range from 700--1300 m for most stations. Typical horizontal separations are of order 1--5 km - in the case of A3, for example, the SPICE borehole is laterally displaced 3230m. A sample A2 VPol waveform used for analysis is shown in Figure \ref{fig:SPICEwf}; in this case, the transmitter was located at a depth of 1077 m.

\subsection{Fitting models to A2 dt(D,R) data}

To assess the fidelity of our n(z) model, we use station A2 dt(D,R) data (selected on the basis of being the most-studied of the current ARA stations) as a `training set' to extract, via $\chi^2$ fit of simulation to data, numerical values for the required one-stage, two-stage, or three-stage n(z) constants (Eqns. (6) and (7)) and then fix those values for comparing data vs. simulations for the other stations. 
For all models, the surface parameter $B_0$ and the asymptotic refractive index 1.778 are fixed. For the multi-stage models, boundary depths $z_1$ and $z_2$
are determined using the SPICE core density data, and then fixed; density boundaries of 550 kg/m$^3$ and 758 kg/m$^3$ imply phase transition boundary depths of 20.5~m and 96.6~m, respectively.  When converted using Eqn. (5), the surface density implies $B_0$=0.45.  


The minimum $\chi^2$ estimate is performed on 368 measured dt(D,R) data points from the SDP across the 8 VPol channels.  The same minimum $\chi^2$ estimate is performed for the 2 stage and 3 stage models with 2 and 3 free parameters for the additional $c_2$ and $c_3$ parameters for the models, respectively.  The minimum $\chi^2$ parameters for the 1 stage, 2 stage, and 3 stage models are presented in Table 1.  
The summed $\chi^2$ per degree of freedom for these models are 406.5, 295.8, and 9.3, respectively.  The higher $\chi^2$ in the 1 stage and 2 stage n(z) model fits to A2 is a result of an inability of the models to fit  dt(D,R) across a range of transmission depths. 

 \begin{figure*}[h]
     \centering
     \includegraphics[width = 0.9\textwidth]{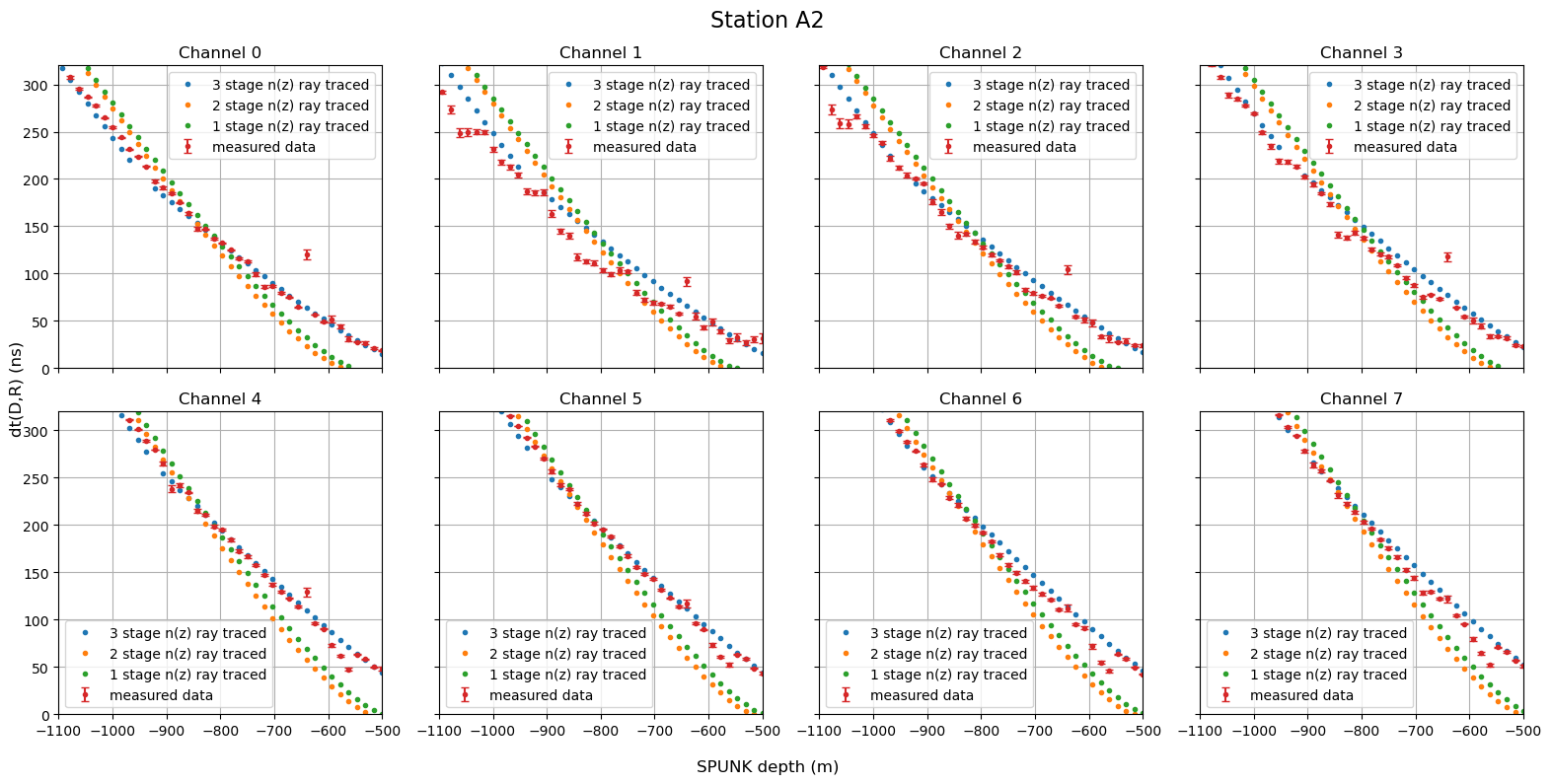}    
     \caption{\small dt(D,R) for Eqn. (5), (6), and (7) compared to measured data for station A2.  Models were fit to minimize $\chi^2$ to data across channels shown, with parameters shown in Table 1.}
    \label{fig:A2fits}
 \end{figure*}
The resulting fitted 1 stage, 2 stage, and 3 stage model parameters are shown in Table 1.  

\begin{table}[ht]
    \centering

\begin{tabular}{| l | l| l | l|} 
  \hline
  n(z) model & $c_1$ & $c_2$ & $c_3$ \\ 
  \hline
  1 stage & 0.0146 & N/A & N/A \\ 
  \hline
  2 stage & 0.0165 & 0.0136 & N/A \\ 
  \hline
  3 stage & 0.0147 & 0.0112 & 0.0310 \\
  
  \hline
\end{tabular}
  \caption{Parameters of A2 dt(D,R) minimum $\chi^2$ estimate for 1 stage, 2 stage, and 3 stage models. 
 }
 \label{tab:model}
\end{table}

\subsection{Comparison to converted SPICE density data}

\begin{figure}[h]
    \centering
    \includegraphics[width = 9.1cm]{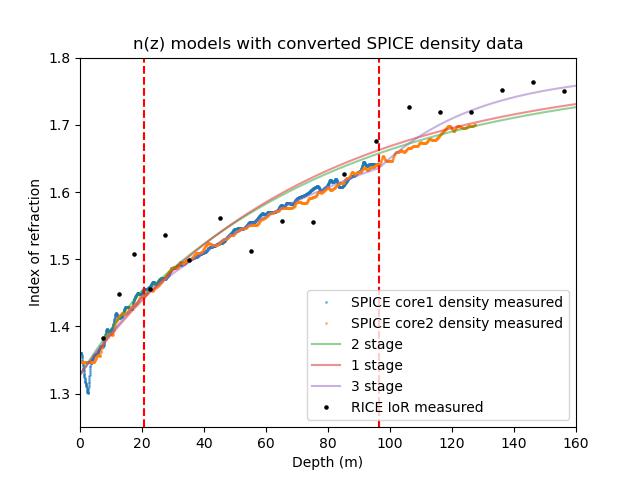}
    \includegraphics[width = 8.6cm]{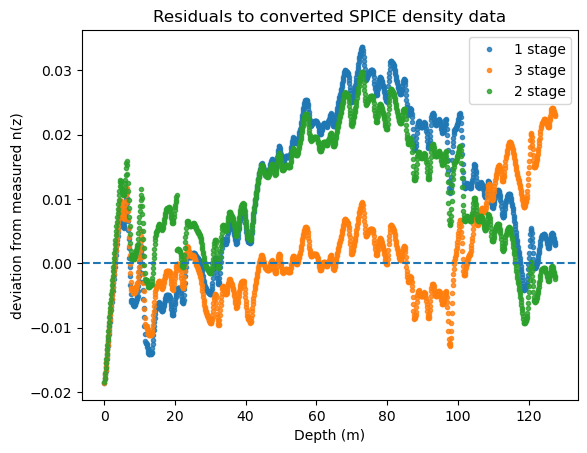}
    \caption{\small  Density profiles corresponding to 1 stage, 2 stage, and 3 stage models (see Table 1) compared to converted SPICE core density data \cite{allison2020long} down to 127m depth.  Deviation between the two SPICE core density measurements suggest an error of $\pm$ 0.005 in measured n(z) values. 
 Residuals shown to converted SPICE core2 density.}
    \label{fig:icemodels_density}
\end{figure}

  We compare the 3 models to SPICE core density data converted to n(z) via Eqn. (5).  For this conversion, we use A=0.845.  Figure \ref{fig:icemodels_density} shows the 1 stage, 2 stage, and 3 stage A2 dt(D,R) parameterizations shown in Table 1.  The residuals show the 3 stage model fitting most closely with the overall SPICE core2 density data, particularly over the firn region between 20.5m and 96.6m.  The root mean squares (RMS) of the deviations for the 1 stage, 2 stage, and 3 stage models to SPICE core2 converted density are 0.0166, 0.0141, and 0.0080 respectively.  However, in the bubbly ice region ($>$ 96.6m), the existing density data shows the 3 stage model exceeding converted n(z).  As the density to refractive index conversion is less studied in this deeper region, it is possible the linear relation breaks down at greater depths.  Additional density data at greater depths would help to draw conclusions about its behavior in this region.  

\subsection{Timing results to other ARA stations}

\begin{figure*}[h]
    \centering    \includegraphics[width=0.85\textwidth]{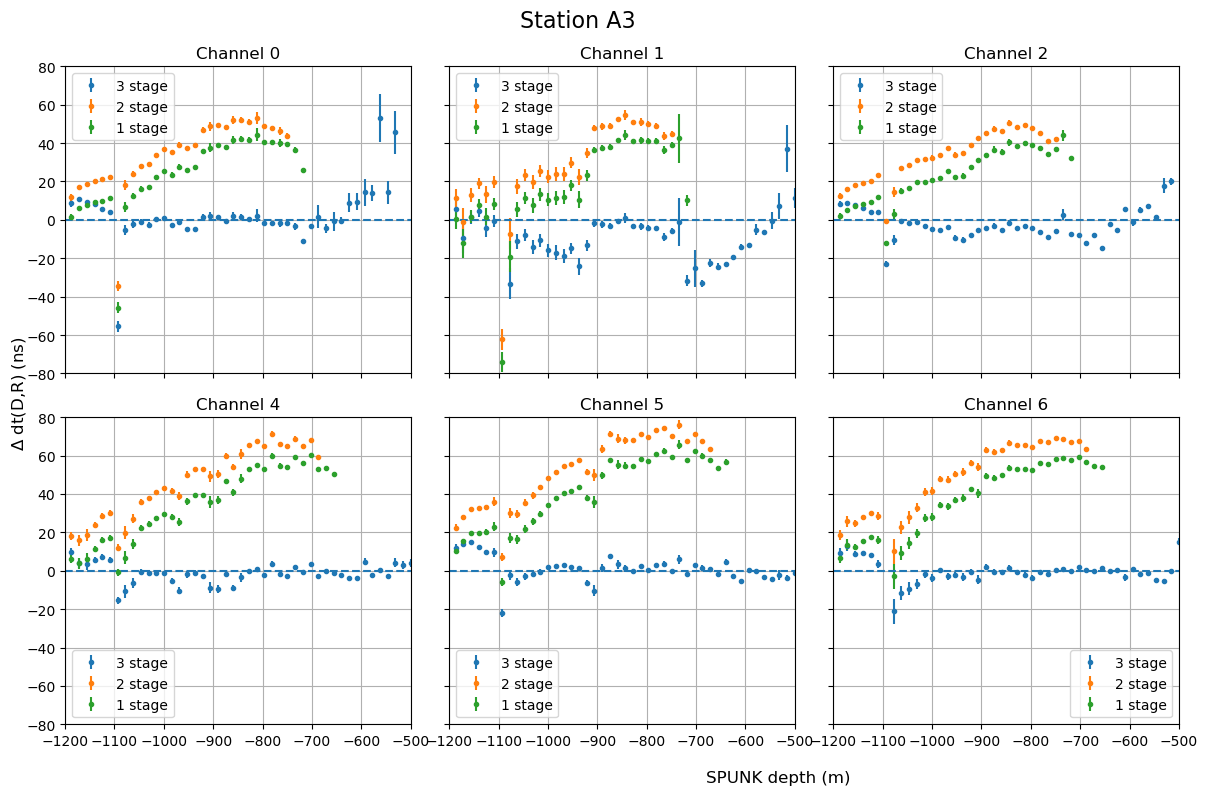}
    \caption{\small $\Delta$ dt(D,R) for signals traveling from SDP to station A3 for Vpol channels.  Shown are comparisons of simulated vs measured dt(D,R) for Eqn. (5), Eqn. (6), and Eqn. (7).}
    \label{fig:A3results}
\end{figure*}

\begin{figure*}[h]
    \centering      \includegraphics[width =0.85\textwidth]{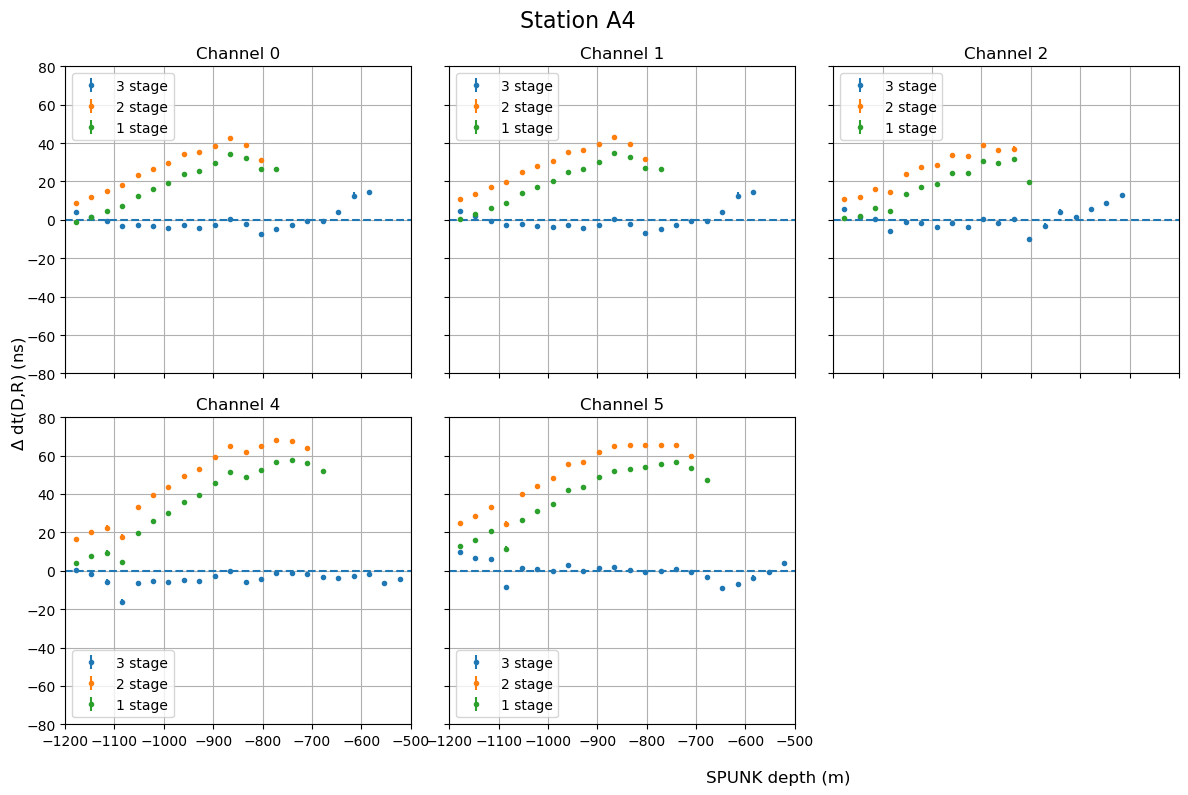}
    \caption{\small $\Delta$ dt(D,R) for signals traveling from SDP to station A4.}
    \label{fig:A4results}
\end{figure*}

\begin{figure*}[h]
    \centering      \includegraphics[width =0.9\textwidth]{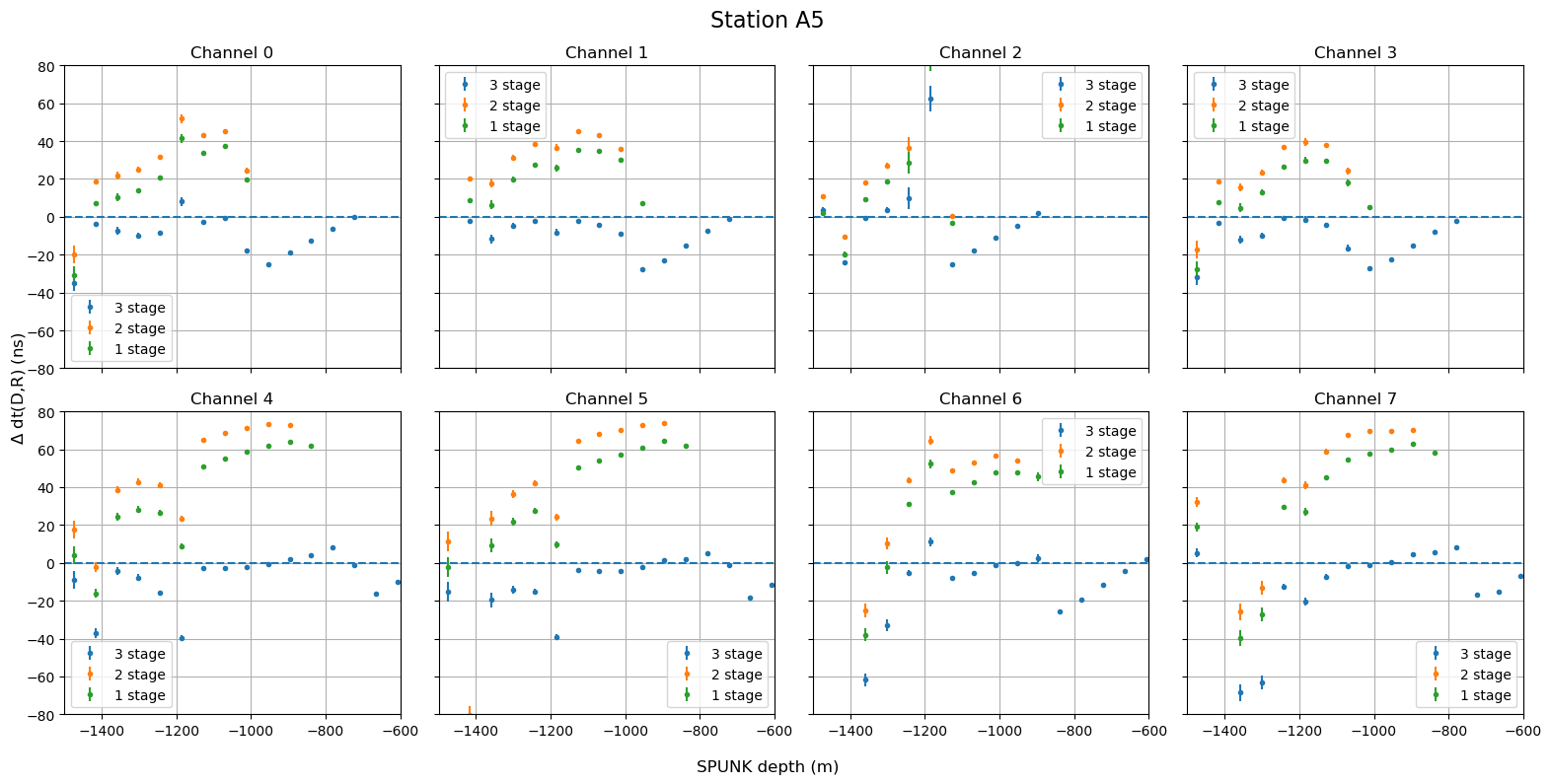}
    \caption{\small $\Delta$ dt(D,R) for signals traveling from SDP to station A5.}
    \label{fig:A5results}
\end{figure*}

Figure \ref{fig:A3results} shows the $\Delta$ dt(D,R) values for various depths of SDP broadcasts to ARA station A3 for the n(z) models corresponding to the 1 stage, 2 stage, and 3 stage parameterizations shown in Table 1.  `Measured' times refer to dt(D,R) calculated using the arrival times of the D and R signals in double pulse waveforms similar to Figure \ref{fig:waveform}, and are determined both by the crossing of an amplitude threshold as well as cross correlation of the D and R signals (see Section III).  
`Simulated' times refer to dt(D,R) calculated from the D and R {\tt RadioPropa}-prescribed ray paths, for given source and receiver locations.  Shown in Figures \ref{fig:A3results} through \ref{fig:A5results} are $\Delta$dt(D,R) values as a function of SDP pulsing depth. Uncertainties in the SDP and receiver antenna coordinates on the (D,R) time differences are assessed by varying these positions in simulations, and incorporated into the $\Delta$dt(D,R) error bars.  

As illustrated in Figures \ref{fig:A3results} and \ref{fig:A4results}, the 1 stage and 2 stage models show a trend for which measured dt(D,R) increases relative to simulated dt(D,R) at shallower depths. The discrepancy can be markedly reduced by using a 3 stage model, for which the $c_3$ parameter is increased in the bubbly ice region relative to the $c$ parameter of the single exponential. 


Station A5 is located 4165m from the SPICE borehole, approximately 1km further away than the other stations.  This larger lateral distance also results in an increased extent of the shadowed zone (see section VII), corresponding to dt(D,R) data only being measurable over a range of 850-1300m source transmitter depth.
Figure \ref{fig:A5results} shows the measured -- simulated dt(D,R) results, comparing the different refractive index parameterizations for station A5. As with the A3 and A4 datasets, we use the $c_3$ parameter determined from A2 data and apply that value to the independent A5 dataset.  The 1 stage and 2 stage models deviate further from the A5 data, which is improved using the 3 stage n(z) model.  The 1 stage and 2 stage models also imply a shadow zone boundary deviating from measured data as there are measured dt(D,R) data points that would be simulated to be in the shadow zone using the 1 stage and 2 stage models.

\section{Shadowed Zone}
As a source moves further away laterally from a receiver or upwards to a shallower depth, dt(D,R) decreases.  Eventually, as dt(D,R) approaches 0, the D and R signals seen in Figure \ref{fig:waveform} begin to overlap.  Initially, this results in focusing that increases signal amplitude.  However, beyond a certain point, corresponding to the shadow zone boundary, the bending of possible paths no longer allows signal to reach the receiver from the transmitter. An example of this would be a transmitter that lies above the blue dashed line shown in Figure \ref{fig:raytrace_path} relative to the receiver. Since refraction is determined by the n(z) model, the shadowed zone, as well as the detected neutrino rate, both depend on n(z).   

\begin{figure}[ht!]
    \centering
    \includegraphics[width = 8.6cm]{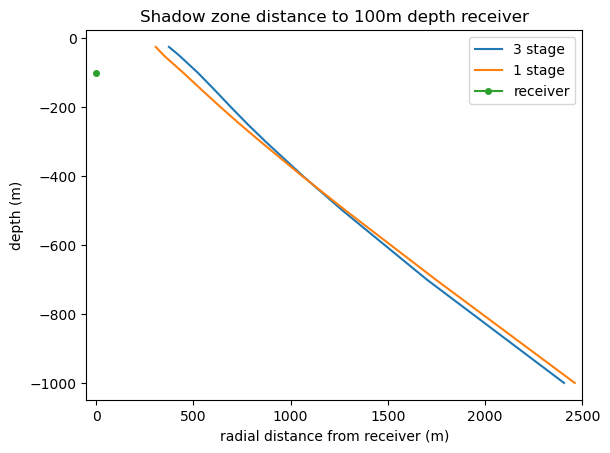}
    \includegraphics[width = 8.6cm]{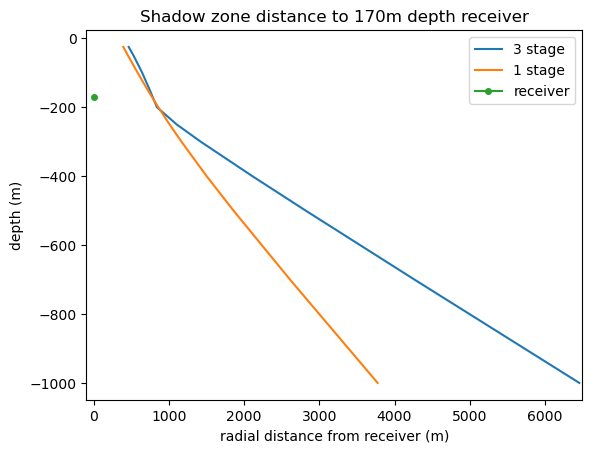}
    \caption{\small Example simulated shadow zone boundary (r,z) coordinates for 100m and 170m depth receiver antennas, for 1 stage and 3 stage n(z) profiles (see Table 1).  Over the range of the SDP pulsing runs, antenna depths exceeding 100m, the 3 stage model implies a smaller shadowed volume, while the 1 stage model implies a smaller shadowed value for 170m depth typical of ARA station antennas.  The 3 stage model yields a greater shadow zone range for sources shallower than 200m depth.}
    \label{fig:shadow_zone}
\end{figure}

Figure \ref{fig:shadow_zone} illustrates how changes in the refractive index model affect the lateral extent of the shadow zone.  Depending on source depth, the 3 stage model changes the distance to the shadow zone boundary for both a 100m and 170m depth receiver antenna, typical of antennas in the RNO-G and ARA experiments, respectively.  Relative to the 1 stage model, the 3 stage model results in a reduction in the distance of the shadow zone boundary for a 100m antenna and an increase of the distance for a 170m antenna for South Polar ice over a distance of source depths from 400m to 1000m.  For source depths shallower than 200m, the 1 stage model increases the distance to the shadow zone boundary. Much of the differences in the shadow zone between the two models results from the differing behaviors of the models in the firn and bubbly ice region.  If the $c_3$ parameter of the 3 stage model were to match the 1 stage model the two shadow zone boundaries are similar, regardless of the differences in the c1 and c2 parameters.  


\section{Impact on neutrino detection}
\begin{figure}[ht!]
    \centering
    \includegraphics[width = 8.6cm]{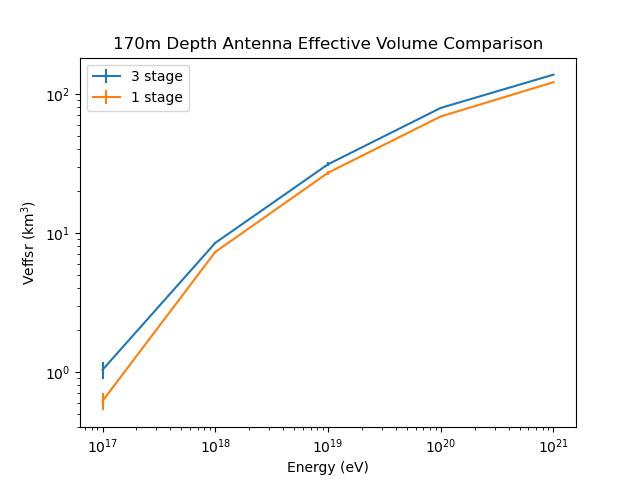}
    \includegraphics[width = 8.6cm]{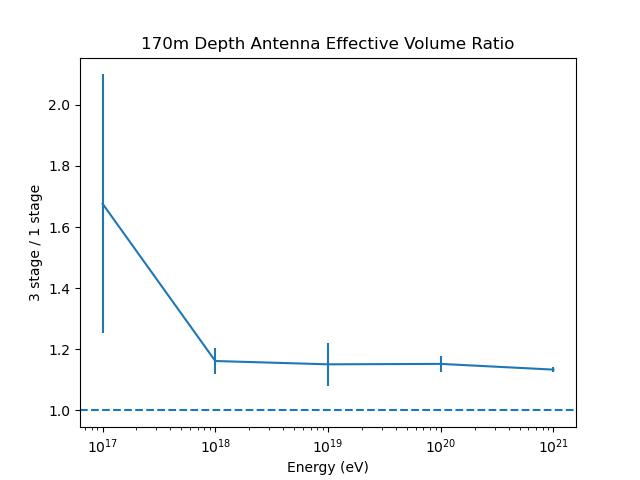}
    \caption{\small Simulated effective volume (top) and ratio (bottom) plots for refractive index profiles specified by equations (5) and (6), for neutrino energies of $10^{17}$, $10^{18}$, $10^{19}$, $10^{20}$, $10^{21}$ eV for 170m depth detector typical of ARA station.  Error bars shown are statistical only.}
    \label{fig:simVeff1}
\end{figure}

\begin{figure}[ht!]
    \centering
    \includegraphics[width = 8.6cm]{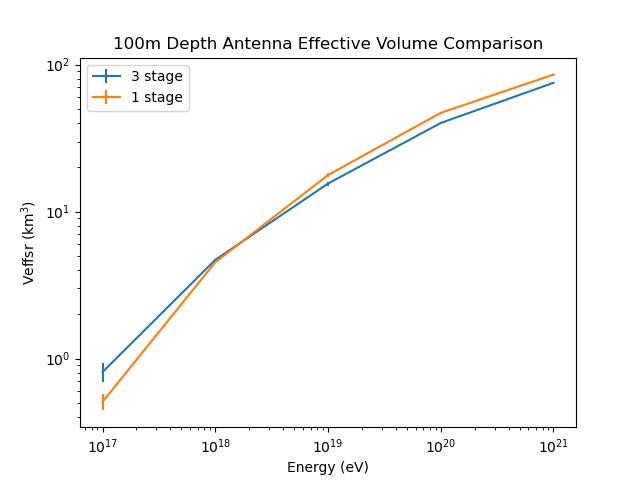}
    \includegraphics[width = 8.6cm]{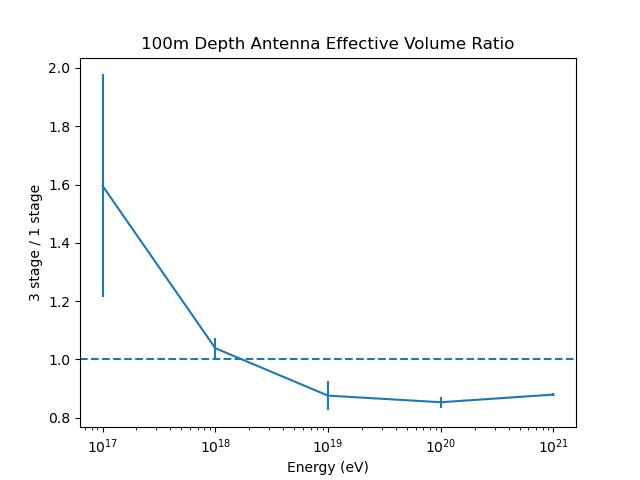}
    \caption{\small Simulated effective volume (top) and ratio (bottom) plots for 100m detector depth.}
    \label{fig:simVeff2}
\end{figure}

Overall, we observe that
the 3 stage model therefore results in a more restricted accessible target volume for 100m depth receiver deployments and correspondingly slightly decreased neutrino sensitivity.
Figures \ref{fig:simVeff1} and \ref{fig:simVeff2} compare effective volume simulations using the two ice models for neutrino detector stations with 170m and 100m deep antenna deployments.  The 3 stage model yields a larger effective volume than the 1 stage model at typical ARA station detector depths.  However, the comparison reverses for more shallow antenna depths typical of other detectors. 

\section{Conclusion}
Measured dt(D,R) time differences from deep radio-frequency transmitters support a glaciologically-motivated 3 stage exponential n(z) model over 1 stage exponential model.  The SPICE core density measurements give no clear determination between the models.  The 3 stage model is closer to converted density in the firn region, but exceeds converted density in the bubbly ice region, where the 1 stage model shows improvement.   Additional density data in the bubbly ice region could provide clarity and constrain the model further.    
Future analysis of the amplitude and frequency content of D and R signals can help refine the n(z) model, as phenomena such as flux focusing are also sensitive to ray curvature.  A comparison of simulated shadowed zone boundaries with those extrapolated from signal amplitudes as a function of depth can also provide an independent check on the refractive index profile.  

An improved n(z) model should also help to provide a more accurate effective volume estimation and aid in current calibration efforts for UHEN experiments in both Greenland and the South Pole, as well as future planned experiments, such as the radio component of the IceCube-Gen2 Radio experiment.

\section{Acknowledgments}
Kenny Couberly was the main author of this manuscript and led the analysis discussed. The ARA Collaboration designed, constructed, and now operates the ARA detectors. We are deeply indebted to the KU Physics and Astronomy Machine Shop, and particularly Scott Voigt and Mark Stockham, who designed and constructed the custom antenna used for recording the primary data essential to this measurement. We would like to thank IceCube and speciﬁcally the winterovers for the support in operating the detector; we also express our appreciation for the authors of the {\tt NuRadioMC} code that was used for our simulations. Data processing and calibration, Monte Carlo simulations of the detector and of theoretical models and data analyses are performed by a large number of collaboration members, who also discussed and approved the scientiﬁc results presented here. We thank the Raytheon Polar Services Corporation, Lockheed Martin, and the Antarctic Support Contractor for ﬁeld support and enabling our work on the harshest continent. We are thankful to the National Science Foundation (NSF) Oﬃce of Polar Programs and
Physics Division for funding support. We further thank the Taiwan National Science Councils Vanguard Program NSC 92-2628-M-002-09 and the Belgian F.R.S.  FNRS Grant 4.4508.01.  
A. Connolly thanks the NSF for Award 1806923 and also acknowledges the Ohio Supercomputer Center. S. A. Wissel thanks the NSF for support through CAREER Award 2033500. M. S. Muzio thanks the NSF for support through MPS-Ascend Postdoctoral Award 2138121.  A. Vieregg thanks the Sloan Foundation and the Research Corporation for Science Advancement, the Research Computing Center and the Kavli Institute for Cosmological Physics at the University of Chicago for the resources they provided. R. Nichol thanks the Leverhulme Trust for their support. K.D. de Vries is supported by European Research Council under the European Unions Horizon research and innovation program (grant agreement 763 No 805486). D. Besson, I. Kravchenko, and D. Seckel thank the NSF for support through the IceCube EPSCoR Initiative (Award ID 2019597)

\clearpage

\bibliography{refs}

\end{document}